\documentclass[aps,pre,twocolumn,groupedaddress,floatfix]{revtex4}

\usepackage{graphicx}
\usepackage{latexsym}

\begin{document}

\begin{widetext}
\noindent\textbf{Preprint of:}\\
D. R. Mason, S. J. Goodman, D. K. Gramotnev, and T. A. Nieminen\\
``Resonant coupling between bulk waves and guided modes in a
dielectric slab with a thick holographic grating''\\
\textit{Applied Optics} \textbf{45}(8), 1804--1811 (2006)
\end{widetext}

\title{Resonant coupling between bulk waves and guided modes\\
in a dielectric slab with a thick holographic grating} 

\author{D. R. Mason}
\author{S. J. Goodman}
\author{D. K. Gramotnev}

\affiliation{Applied Optics Program,
School of Physical and Chemical Sciences,
Queensland University of Technology,
GPO Box 2434, Brisbane, Queensland 4001, Australia}

\author{T. A. Nieminen}

\affiliation{Centre for Biophotonics and Laser Science, Department of Physics,
The University of Queensland, Brisbane QLD 4072, Australia}

\begin{abstract}
What we believe to be a new type of resonant coupling of an incident
bulk wave into guided modes of a slab with a thick holographic grating
is shown to occur in the presence of strong frequency detunings of the
Bragg condition. This happens through the reflection of the strongly
noneigen $+1$ diffracted order with the slab--grating boundaries,
the resultant reflected waves forming a guided slab mode. Rigorous
coupled-wave analysis is used for the numerical analysis of the
predicted resonant effects. Possible applications include enhanced
options for the design of multiplexing and demultiplexing systems,
optical signal-processing devices, optical sensors,
and measurement techniques.
\end{abstract}

\maketitle 

\section{Introduction}

Resonances and anomalies of scattering of electromagnetic waves in
periodic gratings have been extensively analyzed in the
past~\cite{ref1,ref2,ref3}. This includes resonant generation of
guided and surface waves by means of bulk waves interacting with
periodic surface-relief gratings~\cite{ref1,ref2}. The associated
anomalies of reflection and the strong resonant increase of the
scattered wave near the interface were called Wood's
anomalies~\cite{ref2,ref3}. They occur when the scattered wave
propagates almost parallel to the interface of the two media
with a grating. The recent rapid development of photonics, optical
communications, and signal processing has resulted in a strong
renewed interest in these anomalies, mainly due to the possibility
of using them for the design of multiplexing and demultiplexing
devices and coupling of bulk electromagnetic waves to guided or
surface modes~\cite{ref4}. All these effects usually take place
in thin gratings of thickness that is smaller than or of the order
of the wavelength (e.g., surface-relief gratings with small amplitude).

Recently, several new and unusual resonant effects in thick gratings
(of thickness that is much larger than the wavelength) have been
predicted theoretically in the geometry of extremely asymmetrical
scattering and grazing-angle scattering (GAS), i.e., when the
scattered wave propagates parallel5 or almost parallel to the
grating boundaries~\cite{ref6}. In particular, strong GAS resonances
have been associated with the generation of a new type of mode in
gratings and photonic crystals---grating
eigenmodes~\cite{ref7,ref8,ref9,ref10}. These modes can be guided
by a slanted grating alone without conventional guiding effects
in the structure~\cite{ref7,ref8,ref9}. At the same time,
increasing the mean dielectric permittivity inside the grating
region, i.e., assuming that the grating is formed in a guiding
slab, results in a number of new modes propagating in the slab~\cite{ref10}.
These modes have nothing to do with the conventional guided slab
modes---they are generated at wrong angles of scattering and have an
unusual field structure inside the slab~\cite{ref10}.

Grating eigenmodes are formed by all the diffracted orders interacting
with each other inside the grating, and thus cannot exist in the absence
of the grating~\cite{ref7,ref8,ref9,ref10}. They appear to be weakly
coupled to the bulk incident wave, which results in strong resonances
of scattering in the grating at angles that do not correspond to
generation of the conventional guided modes~\cite{ref11}.
In fact, grating eigenmodes can exist only if the conventional
guided modes are not effectively generated due to large grating
amplitude and or grating width, and, vice versa, structures in which
conventional guided modes are effectively generated cannot support
grating eigenmodes. Therefore grating eigenmodes exist only in
sufficiently wide gratings with large amplitude
(e.g., in photonic crystals).

Remarkably, grating eigenmodes can exist (and the associated resonances
can be very strong) in the presence of strong frequency detunings of
the Bragg condition~\cite{ref9,ref12}. In fact, in some structures,
the strongest resonances occur when the frequency is almost halfway
between the two Bragg frequencies corresponding to the first- and
second-order scattering~\cite{ref9}. However, strong resonances of
scattering at frequencies that are strongly detuned from the Bragg
frequency normally occur at large grating amplitude (more than
$\approx 10$\% of the mean permittivity in the grating)~\cite{ref9}.
If the grating amplitude is reduced, the resonances disappear and
eventually are replaced by resonances caused by the generation of
conventional slab modes~\cite{ref10}.

The aim of this paper is to investigate theoretically a new
effect that is associated with strong and sharp resonances
in wide holographic gratings in a guiding slab with small
grating amplitude and at large (up to $\approx 50$\%) detunings
of the Bragg frequency. Remarkably, it will be shown that the
scattered wave is not only strongly noneigen in the slab, but
also propagates in a direction that does not correspond to a
guided mode of the slab. Nevertheless, the predicted resonances
will be explained by generation of the conventional guided modes,
but in a peculiar way by means of interaction of the noneigen
(due to strong frequency detunings) scattered wave from the
slab boundaries.

\section{Structure and methods}

Consider a slanted holographic grating in a thick dielectric slab of
width $L \gg \lambda$ (the wavelength in vacuum) with the dielectric
permittivity given as
\begin{equation}
\epsilon(x,y) =
\begin{array}{ll}
\epsilon_1 & \textrm{for}\,\, x<0,\\
\epsilon_2 + \epsilon_g \exp(\mathrm{i}\mathbf{q}\cdot\mathbf{r}) +
\epsilon_g^\ast \exp(\mathrm{i}\mathbf{q}\cdot\mathbf{r}) &
\textrm{for}\,\,0<x<L,\\
\epsilon_3 & \textrm{for}\,\, x>L
\end{array}
\end{equation}
where $\epsilon_2$ is the mean permittivity in the grating region;
$\epsilon_1$ and $\epsilon_3$ are the permittivities in front of and
behind the grating, respectively;
$\Delta\epsilon_1 = \epsilon_1 - \epsilon_2 < 0$ and
$\Delta\epsilon_2 = \epsilon_3 - \epsilon_2 < 0$ (guiding slab);
$\epsilon_g$ is the grating amplitude; $\mathbf{q} = (q_x,q_y)$
is the reciprocal lattice vector; and $q = 2\pi/Lambda$,
with $\Lambda$ as the grating period. The system of coordinates is
presented in figure 1; all the media are isotropic and nondissipative,
and the grating is infinite in the $y$ and $z$ directions.

\begin{figure}[!htbp]
\centerline{\includegraphics[width=0.8\columnwidth]{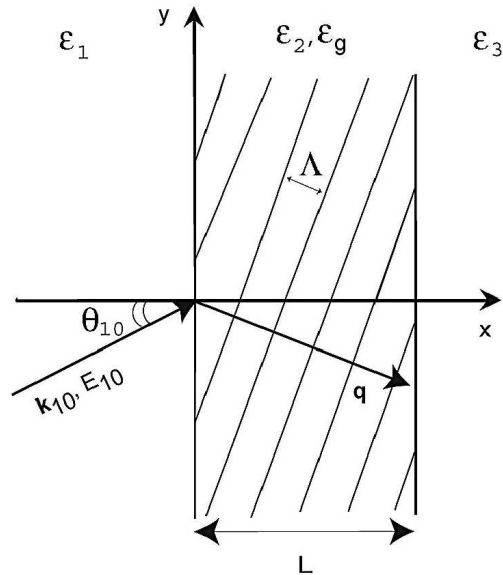}}
\caption{Thick holographic grating with period $\Lambda$, amplitude
$\epsilon_g$, and reciprocal lattice vector $\mathbf{q}$ ($q= 2\pi/\Lambda$)
in a guiding dielectric slab of width $L$. All the materials are assumed
to be linear, isotropic, and lossless; the dielectric permittivities in
front and behind the guiding slab are $\epsilon_1$ and $\epsilon_3$,
respectively. The mean permittivity in the slab is $\epsilon_2$.
A bulk TE electromagnetic plane wave with wave vector $\mathbf{k}_{10}$
and amplitude $E_{10}$ is incident onto the slab at the angle
$\theta_{10}$ with respect to the $x$ axis. The axes of coordinates
are also presented.}
\end{figure}

Let a TE electromagnetic wave with amplitude $E_{10}$ and wave vector
$\mathbf{k}_{10}$ ($k_{10} = \epsilon_1^{1/2}\omega/c$, where
$\omega$ is the angular frequency and $c$ is the speed of light in vacuum)
be incident onto the grating from the left ($x<0$) at an angle
$\theta_{10}$ with respect to the $x$ axis (figure 1).
In the rigorous coupled-wave theory, the field $E(x,y)$ inside the
grating is represented in the form of a
superposition of an infinite number of spatial harmonics
(diffracted orders) with $x$-dependent amplitudes $E_{2n}(x)$~\cite{ref4}:
\begin{equation}
E(x,y) = \sum_{n=-\infty}^{+\infty} E_{2n}(x)
\exp(\mathrm{i}k_{2nx} x + \mathrm{i}k_{2ny} y - \mathrm{i}\omega t),
\end{equation}
and the wave vectors $\mathbf{k}_{2n}$ are determined by the Floquet
condition~\cite{ref4}:
\begin{equation}
\mathbf{k}_{2n} = \mathbf{k}_{20} - n\mathbf{q},
\end{equation}
where $\mathbf{k}_{20}$ is the wave vector of the incident wave
(the 0th diffracted order) in the grating (slab), with
$k_{20} = \epsilon_2^{1/2} \omega / c$.

Substituting equation (2) into the wave equation in the grating
gives an infinite set of coupled-wave secondorder differential
equations that are then truncated and solved numerically for the
amplitudes of the diffracted orders $E_{2n}(x)$ (see, for example,
\cite{ref4}). The unknown constants of integration are determined
from the boundary conditions at the grating boundaries. If at some
frequency $\omega = \omega_0$ and $n=1$ we have $k_{21} = k_{20}
=  \epsilon_2^{1/2} \omega_0 / c$, then the Bragg condition is
satisfied for the $+1$ diffracted order, and $\omega_0$ is called
the Bragg frequency (first-order Bragg scattering). In this case,
the $+1$ diffracted order is an eigen bulk wave in the material of
the slab. If the frequency $\omega$ is such that $k_{21} \ne k_{20}$,
then the Bragg condition is detuned by the value
$\Delta\omega = \omega - \omega_0$.

In this paper we will investigate the frequency response of thick
holographic gratings (described by equation (1)) in a guiding slab.
The grating amplitude is assumed to be small, so that the GAS
resonances do not occur, and resonant generation of conventional
slab modes is expected in the slab. In Section 3, strong anomalies
of scattering will be predicted and explained at large (up to
$\approx 50$\%) negative frequency detunings of the Bragg condition,
i.e., when the frequency of the wave $\omega \approx \omega_0/2$.
Rigorous coupled-wave theory based on the enhanced \textit{T}-matrix
algorithm~\cite{ref13} will be used for the analysis, although in
some cases the approximate coupled-wave theory based on the two-wave
approximation~\cite{ref14} will be sufficient.

\section{Numerical results}

As an example, consider a structure with the following parameters:
$L=10\,\mu$m; $\epsilon_2 = 5$; $\epsilon_g = 2\times 10^{-3}$;
$\theta_{10} = 45^\circ$; $\epsilon_1 = \epsilon_3 = 4.8492$,
i.e., $\Delta\epsilon_1 = \Delta\epsilon_2 = -0.1508$ (the value of
$\Delta\epsilon$ has been chosen so that the critical angle of total
internal reflection for a wave propagating inside the layer is equal
to $80^\circ$). The Bragg condition is assumed to be satisfied
precisely at the frequency $\omega_0 = 1.88\times 10^{15}$ rad/s
(the corresponding Bragg wavelength in vacuum is $\lambda_0 = 1\,\mu$m).

\begin{figure}[!b]
\centerline{\includegraphics[width=\columnwidth]{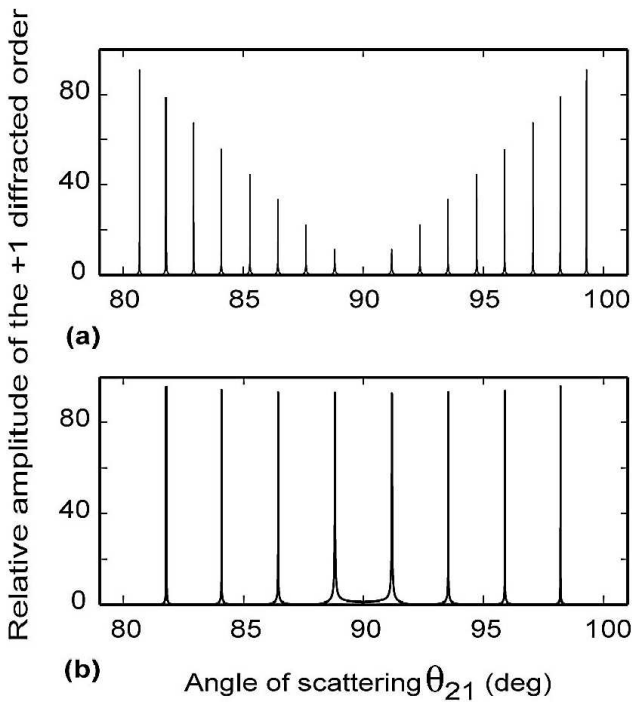}}
\caption{Dependencies of the relative amplitudes of the $+1$
diffracted order $|E_{21}/E_{10}|$ on the angle of scattering
$\theta_{21}$ (i.e., angle between the vector $\mathbf{k}_{21}$
and the $x$ axis in the grating): (a) at the front and rear grating
boundaries (i.e., at $x = 0$ and $x=L$) and (b) in the middle of the
grating (i.e., at $x=L/2$). The slanting angle for grating fringes
and the period $\Lambda$ are adjusted for each angle of scattering
$\theta_{21}$ so that the $+1$ diffracted order satisfies the Bragg
condition precisely (i.e., $k_{21} =  \epsilon_2^{1/2} \omega_0 / c$).
The structural and wave parameters are $L=10\,\mu$m,
$\lambda_0$ (vacuum) $= 1\,\mu$m, $\epsilon_2 = 5$,
$\epsilon_1 = \epsilon_3 = 4.8492$ (the critical angle in the slab is
$80^\circ$), $\epsilon_g = 2\times 10^{-3}$, and the angle of incidence
$\theta_{10} = 45^\circ$.}
\end{figure}

Figure 2 presents the dependencies of the relative amplitude of the
scattered wave ($+1$ diffracted order) $|E_{21}/E_{10}|$ in the grating
on the angle of scattering $\theta_{21}$. This is the angle of
propagation of the $+1$ diffracted order in the grating with respect
to the $x$ axis.
The Bragg condition is assumed to be satisfied precisely for all
considered angles of scattering, i.e., the slanting angle and the
period of the grating are adjusted accordingly for each angle of
scattering. The dependencies are presented at the front and rear
boundaries of the grating (guiding slab) (figure 2(a)) and in the
middle of the grating (slab) (figure 2(b)).

Sharp resonances of scattering can be seen in figure 2, indicating a
strong increase of the scattered wave amplitude at several resonant
angles of scattering. At these angles, resonant coupling between the
incident bulk wave and the modes guided by the slab takes place. The
maximums in figure 2(b) represent the symmetric modes of the slab.
This is because the electric field of the antisymmetric modes in the
middle of the slab is equal to zero (and we do not see the corresponding
resonances in figure 2(b)). At the same time, at the slab boundaries,
both symmetric and antisymmetric modes have a nonzero electric field.
Therefore twice as many maximums can be seen in figure 2(a), corresponding
to both symmetric and antisymmetric modes.

The patterns of the resonances are almost symmetric with respect to
$\theta_{21} = 90^\circ$ in figures 2(a) and 2(b). This is because a
guided mode in the slab can be represented by a bulk wave successively
reflecting from the slab boundaries~\cite{ref11}. Therefore such a mode
can equally be generated by means of coupling of the incident bulk wave
into the $+1$ diffracted order propagating toward the front grating
boundary ($\theta_{21} > 90^\circ$) or toward the rear grating
boundary ($\theta_{21} < 90^\circ$), as long as this diffracted
order propagates at the angle (with respect to either of the slab
boundaries) corresponding to a guided mode. Therefore the pattern
of resonances is symmetric with respect to $\theta_{21} = 90^\circ$.
As a result, the eight maximums in figure 2(b) correspond to four
different symmetric slab modes (pairs of maximums that are symmetric
with respect to $\theta_{21} = 90^\circ$ correspond to the same mode).
Similarly, 16 maximums in figure 2(a) (at the slab boundaries)
correspond to four symmetric and four antisymmetric modes. Note,
however, that the mentioned symmetry disappears when the thickness
of the slab becomes of the order of or less than the wavelength of
the incident wave~\cite{ref15}. In this case, the Bragg regime of
scattering is replaced by the Raman--Nath regime~\cite{ref4},
and it is hardly possible to speak about a slanting angle of the
grating fringes with respect to the slab boundaries. This effectively
means that the reciprocal lattice vector of the grating is parallel
to the slab, rather than pointing at some angle with respect to the
grating slab boundaries. As a result, the pattern of the resonances
becomes asymmetric with respect to $\theta_{21} = 90^\circ$,
and the maximum(s) at $\theta_{21} > 90^\circ$~\cite{ref15}.

We emphasize again that thick holographic gratings in a guiding
slab resonantly couple the incident radiation into a slab mode
if the angle of propagation of the $+1$ diffracted order (for which
the Bragg condition is satisfied) is equal to the angle of propagation
of the bulk wave representing the guide mode of the slab.

Let us now consider the frequency response of scattering in the
thick holographic gratings in a guiding slab. For example, consider
the same structure with the thick holographic structure in a slab
of thickness $L=10\,\mu$m, with $\epsilon_2 = 5$;
$\epsilon_g = 2\times 10^{-3}$; $\theta_{10} = 45^\circ$;
$\epsilon_1 = \epsilon_3 = 4.8492$,
i.e., $\Delta\epsilon_1 = \Delta\epsilon_2 = \Delta\epsilon = -0.1508$.
However, this time we fix the orientation and the period of the grating
so that the Bragg condition for the $+1$ diffracted order is satisfied
precisely at $\omega = \omega_0$ and the angle of scattering is
$\theta_{21} = 70^\circ$. At the mentioned structural parameters,
we change the frequency of the incident wave. The dependencies of
the amplitude of the $+1$ diffracted order on frequency detuning of
the Bragg condition $\Delta\omega$ are presented in figures 3(a)
and 3(b) at the boundaries and in the middle of the slab, respectively.

\begin{figure}[!htbp]
\centerline{\includegraphics[width=\columnwidth]{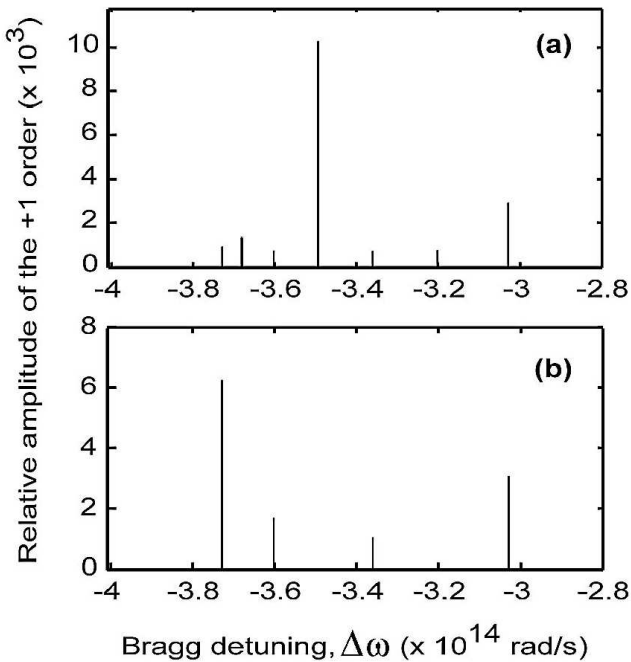}}
\caption{Dependencies of the relative amplitudes of the $+1$
diffracted order $|E_{21}/E_{10}|$ on frequency detuning
$\Delta\omega$: (a) at the front and rear grating boundaries
(i.e., at $x=0$and $x=L$) and (b) in the middle of the grating
(i.e., at $x=L/2$). The structural parameters are the same as
in figure 2. However, this time we fix the grating orientation
and period so that the $+1$ diffracted order satisfies the Bragg
condition at $\theta_{21} = 70^\circ$ and
$\omega = \omega_0 = 1.88\times 10^{15}$ rad/s ($\lambda_0 = 1\,\mu$m).}
\end{figure}

The main unusual features of these dependencies (figure 3) are the
strong and sharp resonances of scattering at the frequency detunings
$\Delta\omega$ that are just $\approx$ 5--6 times smaller than the
Bragg frequency $\omega_0 = 1.88\times 10^{15}$ rad/s. Note that it is
hardly possible to expect any scattering in the grating at the
considered small grating amplitude $\epsilon_g$ and such large
frequency detunings. It is obvious that these resonances must be
related to the resonant generation of some type of localized or
guided modes in the structure. Further decrease of the grating
amplitude results in a rapid increase of the height and sharpness
of the considered resonances. However, the angles at which these
resonances occur are hardly affected by reducing the grating amplitude.
If the grating amplitude tends to zero, then the resonances tend to
infinity, which means that the corresponding modes still exist, but
they become uncoupled to the incident wave. This
demonstrates that these modes can exist in the absence of the grating.
Therefore they cannot be a type of grating
eigenmode~\cite{ref7,ref8,ref9,ref10} but must simply be the
conventional guided modes of the dielectric slab. This has also
been confirmed by considering the field distributions in the grating
and slab at the resonant frequency detunings. It can be shown that
these distributions are practically identical to those corresponding
to the conventional guided modes in the slab. As a result, it can be
seen that the resonances in figure 3(b) correspond to the generation
of the symmetric modes of the slab, and the additional maximums in
figure 3(a) correspond to the antisymmetric modes. In particular,
the leftmost peaks in figures 3(a) and 3(b) (at the largest $\Delta\omega$)
correspond to the fundamental guided mode of the slab, and the peaks
at smaller detunings represent the higher-order guided modes.

However, there is still a question as to how these slab modes could
be generated by the incident wave at the mentioned strong (up to
$\approx 20$\%) frequency detunings $\Delta\omega$ of the Bragg
condition, when the $+1$ diffracted order in the grating is expected
to be a strongly noneigenwave with the wrong wave vector
$\mathbf{k}_{21} = \mathbf{k}_{20} - \mathbf{q}$,
$k_{21} \ne \epsilon_2^{1/2}\omega / c$ ($\omega =\omega_0 +\Delta\omega$),
which does not correspond to a bulk wave propagating in the material of
the slab. Moreover, it can be shown that at each of the resonant
detunings (figure 3) the angle between the vector $\mathbf{k}_{21}$
and the slab boundaries is significantly different from the angle that
one would expect for a guided slab mode. Therefore noneigen
oscillations of the electromagnetic field corresponding to the
$+1$ order should be driven by the incident wave in the grating
and must rapidly tend to zero if $\epsilon_g \rightarrow 0$.
Such a noneigenwave in the slab cannot form a guided mode. This,
however, is in obvious contradiction with the obtained rigorous
numerical results (figure 3).

\begin{figure}[!htbp]
\centerline{\includegraphics[width=\columnwidth]{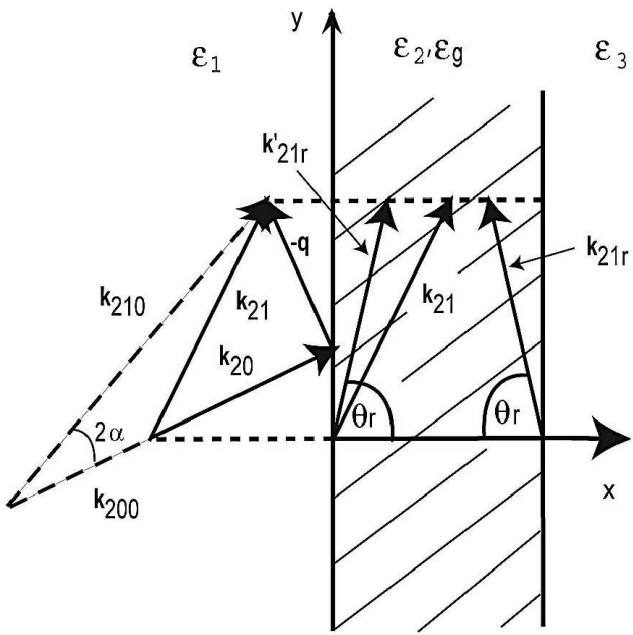}}
\caption{Scheme of coupling between the bulk waves and slab modes by
means of the interaction of the noneigen $+1$ diffracted order with
the grating boundaries: $\mathbf{k}_{200}$ and $\mathbf{k}_{210}$
are the wave vectors of the 0th and $+1$ orders in the slab when the
Bragg condition is satisfied ($\Delta\omega = 0$ and
$\omega =\omega_0$); $\mathbf{k}_{20}$ and $\mathbf{k}_{21}$ are the
wave vectors of the 0th and $+1$ orders at nonzero frequency detuning
$\Delta\omega < 0$; $\mathbf{k}_{21r}$ and $\mathbf{k}'_{21r}$
are the wave vectors of the waves resulting from the reflection
of the noneigen $+1$ diffracted order (with the amplitude $E_{21s}$
and the wave vector $\mathbf{k}_{21}$) from the grating boundaries:
$k_{21ry} = k'_{21ry} = k_{21y}$ (Snell's law) and
$k_{21r} = k'_{21r} = (\omega_0 + \Delta\omega)(\epsilon_2)^{1/2}/c$
(i.e., the reflected waves are eigen for the slab). When the angle of
propagation of these reflected waves $\theta_r$ corresponds to one of
the guided slab modes, resonant generation of this mode occurs with a
strong resonant increase of the amplitude of the $+1$ diffracted order
in the grating (figures 3, 5, and 6). This is because the reflected
waves are formally included in the $+1$ diffracted order (see equations
(2) and (5)).}
\end{figure}

To understand the physical origins of the resonances observed in
figure 3, consider figure 4. If $\Delta\omega = 0$ (i.e.,
$\omega =\omega_0$), then the wave vector of the incident wave
inside the grating and slab is $\mathbf{k}_{200}$. In this case,
we assume that the Bragg condition is satisfied and thus the $+1$
diffracted order is a propagating eigenwave in the slab. Thus the
wave vector of this wave $\mathbf{k}_{210} = \mathbf{k}_{200} - \mathbf{q}$
is equal in magnitude to $\mathbf{k}_{200}$, i.e.,
$k_{210} = \epsilon_2^{1/2}\omega_0 / c$ (the angle between the
$\mathbf{k}_{210}$ vector and the $x$ axis is
$\theta_{210} = 70^\circ$---see above).

Introducing nonzero detunings of the Bragg condition by reducing
the frequency of the incident wave by $\Delta\omega$ results in
reduction of the wave vector of the incident wave $\mathbf{k}_{20}$
(figure 4). At all frequency detunings, the wave vector $\mathbf{k}_{21}$
of the $+1$ diffracted order is given by the Floquet condition
(equation (3)) with $n=1$. Therefore it can also be seen that if
$\Delta\omega < 0$, then $|\mathbf{k}_{21}| > |\mathbf{k}_{20}|$,
and the angle between the $\mathbf{k}_{21}$ vector and the $x$ axis
increases with increasing magnitude of the detuning (figure 4).

As mentioned above, the wave vector $\mathbf{k}_{21}$ has a wrong
magnitude ($k_{21} \ne \epsilon_2^{1/2}\omega / c$) and direction
(that does not correspond to a guided mode in the slab). The
noneigenwave with this wave vector interacts with
the slab--grating boundaries and generates a periodic field pattern
along these boundaries. According to the Huygens principle, this
pattern will generate a reflected bulk wave in the slab with the wave
vector $\mathbf{k}_{21r}$ (figure 4). The tangential (to the boundaries)
component of this vector $\mathbf{k}_{21r}$ is equal to the $y$
component of the wave vector $\mathbf{k}_{21}$: $k_{21ry} = k_{21y}$
(Snell's law). However, since this reflected wave is not driven by
scattering (its wave vector does not satisfy the Floquet condition of
equation (3)), it must be either an evanescent wave in the grating
(if $k_{21ry} > \epsilon_2^{1/2}\omega / c$) or a propagating eigenwave
in the material of the slab (if $k_{21ry} \le \epsilon_2^{1/2}\omega / c$).
Therefore the $x$ component of its wave vector $\mathbf{k}_{21r}$
is given as
\begin{equation}
k_{21rx} = - \sqrt{\epsilon_2\omega^2/c^2 - k_{21y}^2} \ne -k_{21x}.
\end{equation}

If this reflected wave is an eigenwave in the slab (i.e.,
$k_{21ry} \le \epsilon_2^{1/2}\omega / c = k_{21r}$) and it propagates
at an angle that corresponds to one of the guided modes in the slab,
then this guided mode will be resonantly generated in the slab (figure 3).
Thus the guided modes of the slab in this case are generated by means of
the interaction of the noneigen $+1$ diffracted order with
the slab boundaries. Only the presence of the boundaries (structural
discontinuities along the $x$ axis) results in transformation of the
wrong wave vector of the $+1$ diffracted order in the presence of
strong frequency detunings into the correct vector corresponding to
a guided slab mode.

Since changing the frequency results in changing the direction of the
$\mathbf{k}_{21}$ vector and its magnitude, the corresponding direction
of the wave vector of the reflected eigenwave in the slab
$\mathbf{k}_{21r}$ will also change. Thus, adjusting frequency detuning,
we automatically adjust the direction and magnitude of the vector
$\mathbf{k}_{21r}$. When the detuning is such that $\mathbf{k}_{21r}$
corresponds to a guided slab mode, resonant generation of such a mode
occurs (figure 3). The resonant detuning must be strong so as to result
in a significant adjustment of the magnitude and direction of the
$\mathbf{k}_{21r}$ vector. Since the detuning is strong, the
amplitude of the noneigen $+1$ diffracted order in the grating is
very small. This means that the coupling of the incident wave into
the corresponding slab mode is very weak, which results in very strong
and sharp resonances in figure 3. Decreasing the grating amplitude
results in a further decrease of the coupling and the corresponding
increase of the height and sharpness of the resonances (figure 3).
However, the resonant frequency detunings will hardly depend on the
grating amplitude, since it does not affect the directions and the
magnitudes of the considered wave vectors (figure 4). This is
clearly confirmed by the numerical results.

The height of the observed resonances (figure 3) is practically
infeasible---it will be hardly possible to achieve such resonances
experimentally because this would require very large relaxation
times (due to weak coupling). This example has been used to highlight
the considered new type of coupling and to demonstrate that the
resulting resonances are associated with the conventional guided
modes generated in an unusual way by means of reflection of the
noneigen $+1$ diffracted order from the slab boundaries. A more
reasonable situation occurs when the grating amplitude is increased
($\epsilon_g = 0.08$) and the corresponding resonances are
significantly lower, as shown in figure 5. As can be seen in figure 5,
the positions of the resonances are practically the same as in
figure 3 with only $\approx 1$\% reduction of the resonant frequency
detunings. This demonstrates very weak dependence of the resonant
detunings on grating amplitude. This is expected, since only
large grating amplitudes are expected to noticeably change the
guiding properties of the slab and the corresponding wave vectors
of the slab modes.

\begin{figure}[!htbp]
\centerline{\includegraphics[width=\columnwidth]{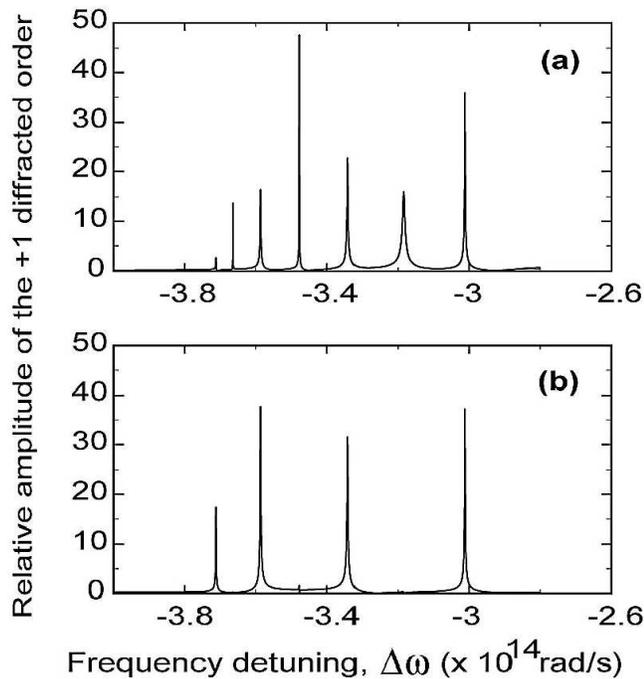}}
\caption{Dependencies of the relative amplitudes of the $+1$ diffracted
order $|E_{21}/E_{10}|$ on frequency detuning $\Delta\omega$: (a) at the
front and rear grating boundaries (i.e., at $x=0$ and $x=L$) and
(b) in the middle of the grating (i.e., at $x=L/2$). The structural and
wave parameters are the same as in figure 3, except for the
increased grating amplitude: $\epsilon_g = 0.08$.}
\end{figure}

The fact that the resonances in figures 5(a) and 5(b) still correspond
to conventional guided modes is again confirmed by the analysis of the
field distribution in the $+1$ diffracted order inside the slab at each
of the resonant frequency detunings. For example, the obtained field
distributions inside that slab for the three leftmost maximums in
figure 5(a) are presented in figure 6(a). It can be seen that these
distributions are very close to those in the first three conventional
modes guided by the slab. Therefore the leftmost maximum in figure 5(a)
corresponds to curve 1 in figure 6(a) (fundamental mode of the slab),
the second-from-the-left maximum in figure 5(a) corresponds to curve 2
in figure 6(a) (the first antisymmetric mode), etc.

\begin{figure}[!htbp]
\centerline{\includegraphics[width=\columnwidth]{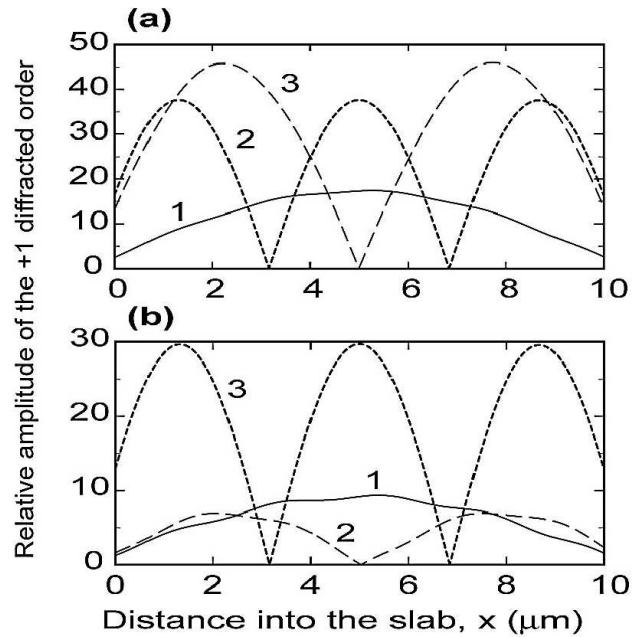}}
\caption{Dependencies on $x$ of the relative amplitudes of the $+1$
diffracted order $|E_{21}/E_{10}|$ inside the grating and slab for the
three leftmost resonances in figure 5(a): (a) $\epsilon_g = 0.08$,
(b) $\epsilon_g = 0.1$. Curves 1, 2, and 3 in (a) correspond to the
leftmost, second leftmost, and third leftmost resonances in figure 5(a).
Resonant detunings are $\Delta\omega_1 \approx -3.71\times 10^{14}$ rad/s,
$\Delta\omega_2 \approx -3.67\times 10^{14}$ rad/s,
$\Delta\omega_3 \approx -3.59\times 10^{14}$ rad/s for curves 1--3,
respectively, in both (a) and (b).}
\end{figure}

At the same time, it can be noticed that curve 1 in figure 6(a)
displays small bumps, which is not a typical feature of the field
distribution corresponding to the fundamental slab mode. These bumps
become more pronounced with increasing grating amplitude (see figure 6(b)
for $\epsilon_g = 0.1$). Curve 2 in figure 6(b) also displays similar
irregularities (unlike curve 2 in figure 6(a) for $\epsilon_g = 0.08$).
This is related to the effect of the grating on the guided modes. This
effect rapidly increases with increasing grating amplitude (compare
figures 6(a) and 6(b)). Eventually, when the grating amplitude becomes
sufficiently large, the structures of the guided modes substantially
change, and they may be transformed into grating
eigenmodes~\cite{ref7,ref8,ref9,ref10}. Note also that the effect
of the grating is more pronounced for lower slab modes (compare the
curves in figures 6(a) and 6(b)).

It is also important to note that the reflected wave with the wave
vector $\mathbf{k}_{21r}$ does not appear explicitly in the coupled-wave
expansion of equation (2). Rather, it is formally included in the
amplitude of the $+1$ diffracted order. Indeed, the $+1$ diffracted order
in equation (2) can be represented as
\begin{eqnarray}
E_(21)(x) \exp(\mathrm{i}\mathbf{k}_{21}\cdot\mathbf{r}
- \mathrm{i}\omega t) = & & \nonumber \\
\left[ E_(21s)(x) + E_(21r) \exp(\mathrm{i}k_{21rx} - \mathrm{i}k_{21x})
\right. & & \nonumber \\
\left. + E'_(21r) \exp(-\mathrm{i}k_{21rx} - \mathrm{i}k_{21x}) \right]
& & \nonumber \\
\times \exp(\mathrm{i}\mathbf{k}_{21}\cdot\mathbf{r} - \mathrm{i}\omega t),
\end{eqnarray}
where $E_{21s}(x)$ is the $x$-dependent amplitude of the $+1$ diffracted
order resulting from the direct interaction (scattering) of the
incident wave with the grating, and the other two terms in the square
brackets are due to the two waves reflected from the rear grating
boundary ($E_{21r}$ with $k_{21rx} < 0$; see equation (4)) and the
front grating boundary ($E'_{21r}$). All three waves have the same $y$
components of their wave vectors (Snell's law, or conservation of the
tangential component of momentum), whereas the $x$ components are
significantly different: $k_{21x}$, $k_{21rx}$, and $-k_{21rx}$,
respectively. Because the $y$ components of the wave vectors are the
same for all three waves, these waves can formally be included in the
$+1$ diffracted order in the expansion in equation (2), which eventually
results in a strong resonant increase of the overall amplitude of this
order at the resonant frequency detunings (figure 3). Actually, these
are the amplitudes $E_{21r}$ and $E'_{21r}$ that experience a strong
resonant increase at the resonant detunings, whereas the amplitude
$E_{21s}(x)$ remains small (it tends to zero as $\epsilon_g \rightarrow 0$).
In this case, because the amplitudes of the reflected waves become
predominant at the resonant detunings, the field distribution inside
the grating and slab indeed appears to be almost identical to that in
the corresponding slab modes (figure 6).

It is possible to derive analytical equations for the resonant detunings
in a slab with a thick holographic grating. For this purpose, we will
need to determine the magnitude and components of the wave vectors
$\mathbf{k}_{21}$ and $\mathbf{k}_{21r}$. From figure 4, it can be seen
that the magnitude of the wave vector $\mathbf{k}_{21}$ is determined
by the equation
\begin{equation}
k_{21}^2 = k_{210}^2 + \Delta k_{20}^2 - 2k_{210}\Delta k_{20} \cos(2\alpha),
\end{equation}
where $\alpha$ is the angle between $\mathbf{k}_{200}$ and the grating
fringes (figure 4) and
\begin{equation}
\Delta k_{20} = k_{200} - k_{20} = -\frac{\Delta\omega}{c}\sqrt{\epsilon_2}.
\end{equation}
It follows that, since $2\alpha$ is the angle between $\mathbf{k}_{200}$
and $\mathbf{k}_{210}$ (figure 4),
\begin{equation}
2\alpha = \theta_{210} - \theta_{200}.
\end{equation}
It can also be seen that the angle $\beta$ between the two vectors
$\mathbf{k}_{21}$ and $\mathbf{k}_{20}$ is determined by the equation
\begin{equation}
q^2 = k_{21}^2 + k_{20}^2 - 2k_{21}k_{20}\cos(\beta).
\end{equation}
Then the angle $\theta_{21}$ between the vector $\mathbf{k}_{21}$ and
the $x$ axis is
\begin{equation}
\theta_{21} = \theta_{200} + \beta.
\end{equation}
Using equations (6)--(10), we find the $y$ component of the wave vector
$\mathbf{k}_{21}$, and thus the components of the $\mathbf{k}_{21r}$
vector:
\begin{equation}
k_{21y} = k_{21}\sin\theta_{21} = k_{21ry},
\end{equation}
\begin{equation}
k_{21rx} = \sqrt{\epsilon_2\omega^2/c^2 - k_{21}^2\sin^2\theta_{21}}
\end{equation}
From here, the angle of propagation of the reflected wave $\theta_r$
(figure 4) is
\begin{equation}
\theta_r = \arctan\left(\frac{k_{21ry}}{k_{21rx}} \right).
\end{equation}
For example, consider the leftmost peak in figure 3. It corresponds
to the frequency detuning $\Delta\omega_1 \approx -3.7279\times 10^{14}$
rad/s. Using equations (6)--(13), we obtain $\theta_r \approx 88.55^\circ$.
The conventional analysis of the guided modes~\cite{ref11} in the
considered slab at the frequency $\omega = \omega_0 + \Delta\omega_1$
demonstrates that this is exactly the angle of the wave propagation
in the slab, which corresponds to the fundamental slab mode. Thus the
leftmost maximum in figure 3 corresponds to the resonant generation
of the fundamental slab mode at the frequency
$\omega = \omega_0 + \Delta\omega_1$ in an unusual way through the
noneigen $+1$ diffracted order, interacting with the slab boundaries.
Using a similar procedure, it can be shown that the other resonances
in figure 3 are associated with the generation of higher slab modes
by means of the same mechanism.

It is important to note that, as can be seen from figure 4, an increase
in the magnitude of the frequency detuning $\Delta\omega$ (i.e., a
reduction of $k_{20}$) results in an increase in the angle $\theta_r$.
This is the reason why the resonance associated with the generation of
the lower-order guided modes (corresponding to larger angles $\theta_r$)
is found at larger detunings (figures 3 and 5).

It is also important to emphasize that the considered
effect of resonant coupling at strong resonant frequency detuning occurs
in a wide range of angles of scattering and incidence. The angle
$\theta_{210} = 70^\circ$ has been chosen just as an example, and a
similar pattern occurs for other scattering angles. However, the actual
values of the resonant frequency detunings will depend on the angles of
scattering and incidence.

Increasing the grating--slab width and/or the mean permittivity in the
slab results in an increase in the number of guided modes. Therefore
this will also result in an increase in the number of different
resonant frequency detunings.

It can be seen that figure 3 displays only seven resonances, whereas
in figure 2 we have eight different maximums corresponding to eight
different modes supported by the slab. It can also be shown that the
number of resonances at strong frequency detunings ($\Delta\omega < 0$)
decreases with decreasing angle of scattering $\theta_1$ and increasing
$|\Delta\omega|$. This is because such strong negative frequency
detunings result in significant reduction of the frequency of the
guided slab modes. Therefore the thickness of the slab is effectively
reduced compared to the wavelength, and the number of modes that can
be sustained by the slab decreases. For example, it can be seen that
this is the highest antisymmetric mode (shown in figure 2) that does
not result in a maximum (i.e., does not exist) in figures 3 and 5.

It can be seen that the considered new way of coupling between an
incident electromagnetic wave and slab modes in the presence of a
thick holographic grating can occur not only by means of the
$+1$ diffracted order but also by means of any other diffraction
order in the grating (e.g., $+2$, $-1$ orders, etc.). In this case,
for example, the strongly noneigen $+2$ diffraction order in the
grating ($k_{22} \ne \epsilon_2^{1/2}\omega /c$) interacts with the
slab boundaries and, if the detuning and the direction of the vector
$\mathbf{k}_{22}$ are appropriate, the resultant reflected wave may
form a guided mode. An equation similar to equation (13) can also be
derived in this case. However, because the higher diffracted orders
are much weaker than the $+1$ order in gratings with small amplitudes
$\epsilon_g$, the efficiency of this coupling will be much lower, and
the corresponding resonances will be very sharp. This makes these
resonances less achievable for the real structures, because they
will correspond to large relaxation times. This problem could be
overcome by increasing grating amplitude, but this will result in
a noticeable effect of the grating on the resultant guided modes.
Therefore the described new type of coupling of bulk waves and guided
slab modes can practically be achieved mainly for the $+1$ diffracted
order (figures 5 and 6), although the principles of this approach can
be applicable to any other higher order of scattering.

\section{Conclusions}

In this paper we have described a new method of coupling an
incident electromagnetic wave into a guided mode of a thick slab
with a holographic grating. As a result, strong resonances were shown to
exist at very large negative frequency detunings of the Bragg condition
in the grating. These resonances rapidly increase with decreasing grating
amplitude, despite the strong (up to $\approx 20$\%) frequency detuning
of the Bragg condition. It was demonstrated that it is the reflection
of the strongly noneigen $+1$ diffracted order from the slab grating
boundaries that may result in a reflected eigenwave that forms a
conventional guided mode of the slab. Thus a new type of resonant
generation of the conventional slab modes in a thick slab with a
holographic grating was demonstrated to occur by means of the
intermediate strongly noneigen $+1$ diffracted order interacting with
the slab--grating boundary.

The predicted resonant coupling of the incident wave into a thick
slab waveguide with a holographic grating may also occur by means
of an $n$th diffracted order (e.g., $n= +2$, $n= -1$, etc.). In this
case, if the grating amplitude is small, the coupling efficiency should
significantly reduce with an increase in the intermediate diffracted
order because the amplitude of this order rapidly decreases with
increasing $n$. This results in a rapid increase of the height and
sharpness of the corresponding resonance with increasing $n$.

Rigorous coupled-wave theory was used for the analysis of
scattering, and the example of bulk TE incident electromagnetic
waves was considered. However, it is important to note that if the
grating amplitude is sufficiently small, the approximate coupled-wave
theory~\cite{ref14} may also be used for the approximate determination
of the predicted resonances at $n = 1$, if reflections at the slab
boundaries are taken into account (see, for example, \cite{ref16}).
At the same time, for other values of $n$, the rigorous coupled-wave
theory is essential even if the grating amplitude is small. This is
because the approximate theory based on the two-wave
approximation~\cite{ref14} simply excludes the higher orders that
are essential intermediaries of the resonant coupling.

Possible applications of the discovered new type of coupling
include enhanced options for multiplexing and demultiplexing,
resonant excitation of slab waveguides, design of new optical
sensors, and measurement techniques.


\begin{thebibliography}{22}

\bibitem{ref1}
V. M. Aranovich and D. L. Mills, \textit{Surface Polaritons:
Electromagnetic Waves at Surfaces and Interfaces} (North-Holland, 1982).

\bibitem{ref2}
R. Petit, \textit{Electromagnetic Theory of Gratings}
(Springer-Verlag, 1980).

\bibitem{ref3}
M. C. Hutley, \textit{Diffraction Gratings} (Academic, 1982).

\bibitem{ref4}
T. K. Gaylord and M. G. Moharam,
Analysis and application of optical diffraction by gratings,
Proc. IEEE 73, 894-938 (1985).

\bibitem{ref5}
D. K. Gramotnev,
Extremely asymmetrical scattering of slab modes in periodic Bragg arrays,
Opt. Lett. 22, 1053-1055 (1997).

\bibitem{ref6}
D. K. Gramotnev,
Grazing angle scattering of electromagnetic waves in periodic Bragg arrays,
Opt. Quantum Electron. 33, 253-288 (2001).

\bibitem{ref7}
D. K. Gramotnev and T. A. Nieminen,
Rigorous analysis of grazing-angle scattering of electromagnetic waves
in periodic gratings,
Opt. Commun. 219, 33-48 (2003).

\bibitem{ref8}
D. F. P. Pile and D. K. Gramotnev,
Second-order grazingangle scattering in uniform wide holographic gratings,
Appl. Phys. B 76, 65-73 (2003).

\bibitem{ref9}
D. K. Gramotnev and D. F. P. Pile,
Frequency response of second-order extremely asymmetrical
scattering in wide uniform holographic gratings,
Appl. Phys. B 77, 663-671 (2003).

\bibitem{ref10}
D. K. Gramotnev, S. J. Goodman, and T. A. Nieminen,
Grazing-angle scattering of electromagnetic waves in gratings
with varying mean parameters: grating eigenmodes,
J. Mod. Opt. 51, 379-397 (2004).

\bibitem{ref11}
A. Yariv and P. Yeh, \textit{Optical Waves in Crystals} (Wiley, 1984).

\bibitem{ref12}
D. K. Gramotnev,
Frequency response of extremely asymmetrical scattering of
electromagnetic waves in periodic gratings,
in \textit{Diffractive Optics and Micro-Optics}, Postconference Digest,
Vol. 41 of OSA Trends in Optics and Photonics
(Optical Society of America, 2000), pp. 165-167.

\bibitem{ref13}
M. G. Moharam, D. A. Pommet, E. B. Grann, and T. K. Gaylord,
Stable implementation of the rigorous coupled-wave analysis for
surface relief gratings: enhanced transmittancematrix approach,
J. Opt. Soc. Am. A 12, 1077-1086 (1995).

\bibitem{ref14}
K. Kogelnik,
Coupled wave theory for thick hologram gratings,
Bell. Syst. Tech. J. 48, 2909-2947 (1969).

\bibitem{ref15}
I. Avrutsky, Department of Electrical and Computer Engineering,
Wayne State University Detroit, Mich. (personal communication, 2004).

\bibitem{ref16}
D. K. Gramotnev, T. A. Nieminen, and T. A. Hopper,
Extremely asymmetrical scattering in gratings with varying mean
structural parameters,
J. Mod. Opt. 49, 1567-1585 (2002).

\end{thebibliography}
\end{document}